\documentclass[twocolumn,prb,10pt,aps,superscriptaddress,floatfix]{revtex4-1}
\usepackage[english]{babel}
\usepackage{graphicx}
\usepackage[pdftex]{epsfig}
\usepackage{ragged2e}
\usepackage{amsmath,amssymb}
\usepackage{times}
\usepackage[colorlinks,linkcolor=blue,citecolor=blue,urlcolor=blue]{hyperref}
\usepackage{color}
\usepackage[table,xcdraw]{xcolor}
\usepackage{epstopdf}
\usepackage{physics}
\usepackage[export]{adjustbox}
\usepackage{feynmf}
\usepackage{dsfont}
\usepackage{bm}

\newcommand{\q}{\bm{q}}
\newcommand{\ri}{\bm{r}_i}

\newcommand{\1}{\hspace*{-1pt}}

\newcommand{\3}{\hspace*{-3pt}}

\begin{document}
\title{Classical spiral spin liquids as a possible route to quantum spin liquids}
\author{Nils Niggemann}
\affiliation{Dahlem Center for Complex Quantum Systems and Institut f\"ur Theoretische Physik, Freie Universit\"{a}t Berlin, Arnimallee 14, 14195 Berlin, Germany}
\author{Max Hering}
\affiliation{Dahlem Center for Complex Quantum Systems and Institut f\"ur Theoretische Physik, Freie Universit\"{a}t Berlin, Arnimallee 14, 14195 Berlin, Germany}
\affiliation{Helmholtz-Zentrum f\"{u}r Materialien und Energie, Hahn-Meitner-Platz 1, 14109 Berlin, Germany}
\author{Johannes Reuther}
\affiliation{Dahlem Center for Complex Quantum Systems and Institut f\"ur Theoretische Physik, Freie Universit\"{a}t Berlin, Arnimallee 14, 14195 Berlin, Germany}
\affiliation{Helmholtz-Zentrum f\"{u}r Materialien und Energie, Hahn-Meitner-Platz 1, 14109 Berlin, Germany}
\date{\today}

\begin{abstract}
Quantum spin liquids are long-range entangled phases whose magnetic correlations are determined by strong quantum fluctuations. While an overarching principle specifying the precise microscopic coupling scenarios for which quantum spin-liquid behavior arises is unknown, it is well-established that they are preferably found in spin systems where the corresponding classical limit of spin magnitudes $S\rightarrow\infty$ exhibits a macroscopic ground state degeneracy, so-called classical spin liquids. Spiral spin liquids represent a special family of classical spin liquids where degenerate manifolds of spin spirals form closed contours or surfaces in momentum space. Here, we investigate the potential of spiral spin liquids to evoke quantum spin-liquid behavior when the spin magnitude is tuned from the classical $S\rightarrow\infty$ limit to the quantum $S=1/2$ case. To this end, we first use the Luttinger-Tisza method to formulate a general scheme which allows one to construct new spiral spin liquids based on bipartite lattices. We apply this approach to the two-dimensional square lattice and the three-dimensional hcp lattice to design classical spiral spin-liquid phases which have not been previously studied. By employing the pseudofermion functional renormalization group (PFFRG) technique we investigate the effects of quantum fluctuations when the classical spins are replaced by quantum $S=1/2$ spins. We indeed find that extended spiral spin-liquid regimes change into paramagnetic quantum phases possibly realizing quantum spin liquids. Remnants of the degenerate spiral surfaces are still discernible in the momentum-resolved susceptibility, even in the quantum $S=1/2$ case. In total, this corroborates the potential of classical spiral spin liquids to induce more complex non-magnetic quantum phases.
\end{abstract}
\maketitle

\section{Introduction}
Quantum spin liquids are prime examples for exotic phases of matter where quantum phenomena and many-body effects combine to create novel emergent properties such as topological order, long-range entanglement and fractional quasiparticle excitations.\cite{anderson73,balents10,savary17} Broadly speaking, quantum spin-liquid behavior may arise in spin systems with sufficiently small magnitudes of the local magnetic moments (typically spin-1/2) when strongly frustrating interactions hinder the system from developing long-range magnetic order in the ground state.

Despite the enormous interest of the condensed matter community in these phases and various decades of intense research activities, quantum spin liquids, however, remain elusive and have so far not been unambiguously detected in real materials. The serious challenges associated with the research on quantum spin liquids also concerns their theoretical understanding; for example, a complete theory describing their correlations, excitations and topological properties does currently not exist. Similarly, a rigorous criterion specifying in which precise microscopic spin models quantum spin liquids exist is also not known. There is, nevertheless, a widespread perception that quantum spin liquids primarily occur in spin systems where the corresponding classical model exhibits a macroscopic ground state degeneracy. A large degeneracy enables stronger thermal and/or quantum fluctuations such that long-range magnetic order may even be suppressed in the classical limit which is known under the term `classical spin liquid'.\cite{balents10,frustration_book}

Previously, the approach of stabilizing quantum spin liquids starting from classical spin liquids has mostly been pursued for kagome\cite{sachdev92,lecheminant92,yan11,han12}, pyrochlore\cite{ramirez99,bramwell01,castelnovo08,gingras14} and related lattices where corner-sharing triangles or tetrahedra give rise to an `ice rule' constraint which is the origin of a residual entropy at zero temperature. A second and less explored possibility of achieving a macroscopic classical ground state degeneracy is via spiral spin liquids.\cite{bergman07,gao16,iqbal18,buessen18,mulder10,henley87,ghosh19,attig17} In these phases, a special interplay of lattice geometries and frustrating long-range interactions induces an exact degeneracy of spiral ground states which is usually of subextensive type, i.e., in two-dimensional systems the wave vectors defining these spirals form closed contours in momentum space while in three dimensions they form surfaces.

Two paradigmatic systems harboring spiral spin liquids are the Heisenberg models on the two-dimensional honeycomb and the three-dimensional diamond lattice with nearest neighbor $J_1$ and antiferromagnetic second neighbor $J_2>0$ couplings.\cite{bergman07,gao16,iqbal18,buessen18,mulder10} Even though these systems are bipartite and, hence, show unfrustrated ferromagnetic or N\'eel magnetic order in the $J_1$-only case, when $J_2>0$ is tuned beyond a certain critical coupling ratio $J_2/|J_1|$, spiral contours or surfaces begin to form marking the onset of a spiral spin liquid. Various numerical studies indeed indicate that both models retain their non-magnetic ground states when replacing the classical $S\rightarrow\infty$ spins by quantum $S=1/2$ spins possibly realizing a quantum spin-liquid phase.\cite{buessen18,reuther11,zhu13,gong13,zhang13,li12} Despite the fragility of spiral spin liquids with respect to order-by-disorder effects and perturbing longer-range couplings which may easily lift the degeneracy,\cite{bergman07,iqbal18,buessen18} an approximate version of this phase has been experimentally identified in the spin-5/2 diamond lattice compound MnSc$_2$S$_4$.\cite{gao16} More recently, experimental and theoretical investigations of the spin-1 diamond spinel NiRh$_2$O$_4$ indicate that this material might realize a situation where quantum spin-liquid behavior originates from a spiral spin liquid in the classical $S\rightarrow\infty$ limit.\cite{buessen18,chamorro18} One reason why this scenario of stabilizing a quantum spin liquid is less explored as compared to the aforementioned kagome and pyrochlore geometries is because there is currently no general criterion known which allows one to systematically construct models with spiral degeneracies.

The purpose of this work is two-fold. Firstly, based on the classical Luttinger-Tisza method,\cite{luttinger46,luttinger51} we formulate an approach to create new models for spiral spin liquids on bipartite lattices. The known spiral spin liquids on the honeycomb\cite{mulder10}, diamond\cite{bergman07} and bcc lattices\cite{attig17} all fall into the category of systems that can be described within this approach. We then extend these considerations and construct two more models harboring spiral spin liquids which are less explored or have not been studied before. The first is a Heisenberg model on the two-dimensional square lattice with antiferromagnetic nearest neighbor $J_1$, second neighbor $J_2$ and third neighbor $J_3$ interactions. Interestingly, even though this system has been extensively studied in the existing literature,\cite{danu16,iqbal16,chubukov91,ferrer93,mambrini06,arlego08,reuther11_2,ralko09} the spiral spin liquid occurring for $J_2/J_1>1/4$ and $J_3/J_2=1/2$ has so far rarely been discussed.\cite{ceccatto93,moreo90} Additionally, we construct a spiral spin liquid on the three-dimensional hcp lattice (hcp stands for `hexagonal close packing') which requires a total of four antiferromagnetic couplings. We show that there is a one-parameter manifold of couplings (i.e. two coupling ratios are fixed) for which the system develops degenerate spiral surfaces in momentum space.

Our second objective is to demonstrate that these spiral spin liquids are good candidate systems for realizing a quantum spin liquid in the $S=1/2$ case. Therefore, we employ the pseudofermion functional renormalization group (PFFRG) method\cite{reuther10,baez18} which is capable of treating strongly frustrated spin systems even in the case of complex two-dimensional\cite{reuther11,reuther10,reuther11_2,baez18,iqbal16,balz16,keles18} or three-dimensional\cite{iqbal19,iqbal18,buessen18,iqbal17,iqbal16_2} lattice geometries and in the presence of long-range couplings.\cite{reuther11,reuther11_2,keles18,iqbal16,iqbal18,buessen18,iqbal17,iqbal16_2} Our numerical results indicate that for both models the coupling regimes of classical spiral degeneracies indeed host extended non-magnetic phases in the $S=1/2$ case. The weight distributions in the static but momentum-resolved spin susceptibilities also show that residual features of the degenerate spiral surfaces still survive in the quantum case, however, depending on the precise coupling parameters the sizes of these surfaces are seen to deviate from their classical values. In total, the two models which we investigate in the classical and in the quantum case constitute additional examples demonstrating that classical spiral spin liquids provide a promising platform for the search for quantum spin liquids.

The remainder of this work is organized as follows: In Sec.~\ref{sec:LT} we investigate under which conditions spiral degeneracies occur on bipartite lattices. These considerations are based on the classical Luttinger-Tisza approach which is briefly introduced in Sec.~\ref{sec:FTransform}. The general scheme for designing spiral spin liquids on bipartite lattices is developed and formulated in Sec.~\ref{sec:Criteria}. The following Sec.~\ref{ch:SSL} makes this procedure more concrete by demonstrating its capability for the square (Sec.~\ref{sub:square}) and for the hcp lattice (Sec.~\ref{sub:hcp}). The effects of quantum fluctuations on both systems are studied in Secs.~\ref{sec:SquareFRG} and \ref{sec:hcpFRG}, respectively, where the PFFRG method is applied to calculate momentum-resolved susceptibilities and to map out phase diagrams. We conclude the paper in Sec.~\ref{sec:conclusion} with a short summary of the main results and an outlook for future directions of research.

\section{Ground state degeneracies in classical bipartite Heisenberg systems}
\label{sec:LT}
In this section we apply the classical Luttinger-Tisza method~\cite{luttinger46,luttinger51} to find a criterion for the formation of spiral spin liquids that can be used for the identification of new systems with this property. Since the Luttinger-Tisza method will be essential for formulating this criterion, we will first give a brief introduction into this approach. The considerations below crucially rely on the fact that our models are defined on bipartite lattices.

\subsection{The Luttinger-Tisza method}
\label{sec:FTransform}
In the classical Heisenberg model, spins are represented as three-dimensional vectors, such that the Hamiltonian of a system with $N$ classical spins $\bm{S}_i = \bm{S}(\ri)$ coupled by exchange interactions $J_{ij}=J(\bm{R}_{ij})$ can be written as
\begin{equation}
  H = \frac{1}{2} \sum_{ij}{J(\bm{R}_{ij})\bm{S}(\ri)\bm{S}(\bm{r}_j)}\;. \label{eq:Heisenberg}
\end{equation}
Here, $\bm{r}_i$ is the real space position of site $i$ and $\bm{R}_{ij} = \ri-\bm{r}_j$ is the distance between two sites $i$ and $j$. Even in the classical case, finding the ground state of Eq.~(\ref{eq:Heisenberg}) is usually a non-trivial minimization problem. The Luttinger-Tisza method solves this problem, at least approximatively, but can also be exact in special cases, as will be discussed in more detail below. This approach starts by defining the Fourier transforms of the spins $\tilde{\bm{S}}_\alpha(\q)$ and the exchange interactions $\tilde{J}_{\alpha \beta}(\q)$ on each of the sublattices $\alpha$,
\begin{align}
\tilde{\bm{S}}_\alpha(\q) &= \frac{1}{\sqrt{N  / \Gamma}} \sum_{i \in \alpha}{\bm{S}(\ri) e^{-i \q\ri}}\;,\\
\tilde{J}_{\alpha \beta}(\q) &= \frac{1}{2} \sum_{j\in\beta}{J(\bm{R}_{i\in\alpha,j})e^{i \q \bm{R}_{i\in\alpha,j}}}\;. \label{eq:LTMatrix}
\end{align}
Here, we a have assumed that the lattice has $\Gamma$ sites per unit cell, i.e., $\alpha=1,\ldots,\Gamma$ runs over all the $\Gamma$ sublattices and $\frac{N}{\Gamma}$ is the number of sites in each sublattice. Furthermore, in the definition of $\tilde{J}_{\alpha \beta}(\q)$ the index $i\in\alpha$ denotes an arbitrary but fixed site on sublattice $\alpha$. Note that $\tilde{J}_{\alpha \beta}(\q)$ can be interpreted as the elements of a $\Gamma\times\Gamma$ matrix $\underline{\bm{J}}(\q)$ which contains all the interactions between the sublattices $\alpha$ and $\beta$. Rewriting the Hamiltonian in Eq.~(\ref{eq:Heisenberg}) in terms of $\tilde{\bm{S}}_\alpha(\q)$ and $\tilde{J}_{\alpha \beta}(\q)$ leads to
\begin{equation}
  H =\sum_{\q}\sum_{\alpha,\beta} \tilde{J}_{\alpha \beta}(\q) \tilde{\bm{S}}_\alpha(\q)\tilde{\bm{S}}_\beta(-\q)\;,
  \label{eq:FTHamilton}
\end{equation}
where the sum $\sum_{\bm{q}}$ runs over all wave vectors in the first Brillouin zone. It is crucial to realize that the matrix $\underline{\bm{J}}(\q)$ is hermitian and, hence, has real eigenvalues. Its normalized eigenvectors form an orthonormal basis of the $\Gamma$-dimensional vector space. Each cartesian component of the Fourier-transformed spin vectors can thus be expressed as a linear combination of eigenvectors $\bm{u}_\nu(\q)$ of $\underline{\bm{J}}(\q)$ and, therefore, we may write
\begin{equation}
\tilde{\bm{S}}_\alpha (\q) = \sum_{\nu = 1}^\Gamma{\bm{w}_\nu(\q) u_\alpha^\nu(\q)} \label{eq:FTConfiguration}
\end{equation}
with $u_\alpha^\nu(\q)$ being the $\alpha$-th component of the $\nu$-th eigenvector and $\bm{w}_\nu(\q)$ the $\nu$-th vector that determines the different cartesian components of $\tilde{\bm{S}}_\alpha (\q)$. Inserting this into Eq.~(\ref{eq:FTHamilton}) results in a Hamiltonian that depends only on the eigenvalues $\lambda_\nu (\q)$ and the coefficients $\bm{w}_\nu(\q)$,
\begin{equation}
H = \sum_{\q }\sum_{\nu = 1}^\Gamma \lambda_\nu(\q) \left|\bm{w}_\nu(\q)\right|^2\;.
\label{eq:EVHamiltonian}
\end{equation}
Additionally, we have the condition that all spins are normalized,
 \begin{equation}
  |\bm{S}_i|^2 = 1 \hspace{10pt} \forall i\;, \label{eq:strongConstraint}
\end{equation}
which is known as the `{strong constraint}'. Minimizing the Fourier-transformed Hamiltonian in Eq.~(\ref{eq:EVHamiltonian}) under the constraint in Eq.~(\ref{eq:strongConstraint}) does not yet simplify the problem. The key conceptual step proposed by Luttinger and Tisza to approximately solve the minimization problem amounts to replacing the `{strong constraint}' by the so-called `{weak constraint}'. The latter is much less restrictive, as it only constrains the total spin, instead of imposing separate conditions for all $N$ particles,
\begin{equation}
  \sum_{i=1}^{N}{|\bm{S}_i|^2} = N\;. \label{eq:WeakConstraint}
\end{equation}
Evidently, all solutions to the problem that satisfy the strong constraint fulfill the weak one as well, whereas the opposite is generally not the case. Therefore, only those solutions found under the weak constraint that additionally satisfy the strong constraint describe physical ground states. In Sec.~\ref{sec:Criteria} it will be shown that under rather mild additional assumptions the Luttinger-Tisza method becomes exact for all bipartite lattices.

Using the inverse Fourier transform of the spins on the $\alpha$-th sublattice,
\begin{equation}
  \bm{S}(\bm{r}_{i\in\alpha}) = \frac{1}{\sqrt{N  / \Gamma}}\sum_{\q}{e^{i \bm{qr}_i} \tilde{\bm{S}}_\alpha}(\q)\;,
\end{equation}
the weak constraint can be re-expressed in Fourier space,
\begin{equation}
  \sum_{\q} \sum_{\nu = 1}^\Gamma {\left|\bm{w}_\nu(\q)\right|^2} = N\;.
  \label{eq:FTWeakConstraint}
\end{equation}
Using Eq.~(\ref{eq:EVHamiltonian}) and Eq.~(\ref{eq:FTWeakConstraint}), a lower limit for the energy follows immediately,
\begin{equation}
  H \geq N \lambda_{LT}\;,
  \label{eq:LowerLimit}
\end{equation}
where the Luttinger-Tisza eigenvalue $\lambda_{LT} \equiv \min{\left\{\lambda_\nu(\q)\right\}}$ is defined as the minimum out of the set of eigenvalues of $\underline{\bm{J}}(\q)$ with respect to all wave vectors. The energy can therefore be minimized by setting the coefficients $\bm{w}_\nu(\q)$ to be only non-zero for the values $\pm \q_{LT}$ and $\nu_{LT}$ that minimize $\lambda_\nu(\q)$,
\begin{equation}
  \left| \bm{w}_\nu (\q)\right|^2 = \begin{cases}
    \frac{N}{2} & \text{if } \q = \pm \q_{LT} \wedge \nu = \nu_{LT}\\
    0 & \text{otherwise}
  \end{cases}\;.
\end{equation}
This can be seen by first noting that from Eq.~(\ref{eq:LTMatrix}) it follows $J_{\alpha \beta}(-\q) = J_{\alpha \beta}^*(\q) = J_{\beta \alpha}(\q)$ and thus $\lambda(\q) = \lambda(-\q)$.
In the case of several equivalent minima at wave vectors $\{\q_{LT}\}$, one may chose non-zero $\bm{w}_\nu(\q)$ for any $\q \in \{\pm \q_{LT}\}$, if the resulting state satisfies the strong constraint. These coefficients by construction satisfy the weak constraint and lead to a spin configuration that is obtained by transforming  Eq.~(\ref{eq:FTConfiguration}) back to real space,
\begin{equation}
    \bm{S}(\bm{r}_{i\in\alpha}) = \frac{1}{\sqrt{2 \Gamma}} \sum_{\q = \pm \bm{q_{LT}}}{\hat{\bm{n}}(\q)
    u_\alpha^{LT}(\q)  e^{i \bm{qr}_i}} \label{eq:SpinConfig}
\end{equation}
Here, $\hat{\bm{n}}(\q)$ is the unit vector along $\bm{w}_{\nu_{LT}}(\q)$ and $u_\alpha^{LT}$ is the $\alpha$-th component of the eigenvector of $\underline{\bm{J}}(\q_{LT})$ that corresponds to the smallest eigenvalue.

Finally, the physical solutions are only those that satisfy the strong constraint in Eq.~(\ref{eq:strongConstraint}). It is now obvious why both $+\q$ and $-\q$ are taken into account: Since the normalized vector $\hat{\bm{n}}(\q)$ is still undetermined, different complex phases for its $x$- and $y$-components must be chosen such that the spins $\bm{S}_i$ are real, which produces coplanar spirals of the form $\bm{S}(\ri)  = (\cos{(\q_{LT}  \ri)} , \sin{(\q_{LT} \ri)}, 0)$. This solution is most evident for mono-atomic Bravais lattices where $u^{LT}(\q) = 1$ and the method becomes exact (at least in the spin-isotropic Heisenberg case). For non-Bravais lattices, the ground state cannot generally be represented by such a spiral state. It should further be stressed that any rotation of this spin configuration is also a possible solution as the isotropic Heisenberg model is invariant under uniform rotations of all spins.

If the Luttinger-Tisza method results in one or more possible spin configurations that fulfill the strong constraint, the physical ground state of the system is found. Vice versa, if no solutions can be obtained that have a normalized  spin configuration, then the method fails, resulting only in a lower limit for the system's energy given by Eq.~(\ref{eq:LowerLimit}). A generalized version of the weak constraint that allows for distinct magnitudes of spins on each sublattice has been proposed by D. H. Lyons and T. A. Kaplan \cite{lyons60}.

\subsection{Criteria for ground state degeneracies on bipartite lattices}
\label{sec:Criteria}
While for Bravais lattices any ground state found via the Luttinger-Tisza method satisfies the strong constraint, this is generally not true in the case of lattices with more than one site per unit cell. In this section, we will analyze the Luttinger-Tisza solution in the special case of bipartite lattices, which consist of two sites per unit cell, i.e., they can be seen as two interpenetrating Bravais lattices with $\alpha\in\{1,2\}$. Particularly, we will show that for equivalent sublattices this solution is still exact. We will then derive a criterion for the coupling parameters such that these solutions exhibit subextensive spiral degeneracies. Our arguments here are formulated in a general way, without specifying the precise lattice geometry. In the next section, we will illustrate these considerations based on two concrete examples.

For any bipartite lattice, frustration is only possible if interactions beyond nearest-neighbor spins are present. Otherwise, neighboring spins from different sublattices can always be aligned or anti-aligned, leading to a ferromagnetic or N\'eel ordered state, respectively. For simplicity, in this section we restrict ourselves to first- and second-neighbor interactions only; the general argument is, however, also valid for further-neighbor couplings. Consider an $n$-dimensional lattice with a set of primitive vectors $\{\bm{a}_i\}$ and a basis $\bm{h}$ defining the relative shift of the two sublattices. For a bipartite lattice, the matrix $\underline{\bm{J}}(\q)$ is of $2 \times 2$ form. Its diagonal elements contain all the interactions between spins that are connected by primitive lattice vectors and, as a result of the inversion symmetry of Bravais lattices, are real,
\begin{equation}
  J_{\alpha \alpha}(\q) = J_2^\alpha \sum_{i=1}^n{\cos{(\q \bm{a}_i })} \equiv J_2^\alpha f(\q)\;. \label{eq:fq}
\end{equation}
The off-diagonal terms ($\alpha \neq \beta$) are possibly complex,

\begin{equation}
  J_{\alpha \beta}(\q)= \frac{J_1}{2} \sum_{\bm{R}^\text{nn}_{\alpha\beta}}{e^{i \q \bm{R}^\text{nn}_{\alpha\beta}}}\;,
  \label{eq:Offdiagonal}
\end{equation}
where the vectors $\bm{R}^\text{nn}_{\alpha\beta}$ point from an arbitrary, but fixed site on sublattice $\alpha$ to all its nearest neighbors within the other sublattice $\beta$. We now make the assumption that the two sublattices are equivalent ($J_2^{\alpha=1} = J_2^{\alpha=2}$) which means that the diagonal elements of the hermitian matrix $\underline{\bm{J}}(\q)$ are equal,
\begin{equation}
  \underline{\bm{J}}(\q) = \begin{pmatrix}
    J_{11}(\q) & J_{12}(\q)\\
    J_{12}^*(\q) & J_{11}(\q)
  \end{pmatrix}\;.
\end{equation}
Diagonalization of this matrix leads to the eigenvectors and eigenvalues
\begin{align}
  \bm{u}_\nu(\q) &= \frac{1}{\sqrt{2}}\begin{pmatrix}
\pm e^{i \phi(\q)}\notag \\
1
\end{pmatrix}\;,
\\
\lambda_\nu(\q) &= J_{11}(\q) \pm |J_{12}(\q)|\;. \label{eq:Eval}
\end{align}
Here, $\phi$ is the angle between the two spins in the same unit cell given by
\begin{equation}
  e^{i \phi (\q)} = \sqrt{\frac{J_{12}(\q)}{J_{12}^*(\q)}}\;. \label{eq:Phi}
\end{equation}
From Eq.~(\ref{eq:SpinConfig}) it follows that the strong constraint can be fulfilled exactly in the case where both components of the Luttinger-Tisza eigenmode $\bm{u}_{\nu_{LT}}(\q_{LT})$ have the same absolute value. Thus, the spin configuration obtained from minimizing the eigenvalues in Eq.~(\ref{eq:Eval}) satisfies the strong constraint for all bipartite lattices with equivalent sublattices such that Luttinger-Tisza becomes exact (note that this property is lost for inequivalent sublattices). As in the case of Bravais lattices, the solution has the form of coplanar spirals,
\begin{equation}
  \bm{S}(\bm{r}_{i\in\alpha})  =
  \begin{pmatrix}
    \cos{(\q_{LT} \ri + \phi_\alpha)}\\
    \sin{(\q_{LT} \ri + \phi_\alpha)}\\
    0
  \end{pmatrix}\;,
\end{equation}
where we use the convention $\phi_1\equiv\phi$ and $\phi_2 = 0$, i.e. $\phi_1$ is the angle between spins in the same unit cell separated by the basis $\bm{h}$.

The remaining question which we address in this subsection is under which conditions the exact solution that follows from Eq.~(\ref{eq:Eval}) exhibits a degenerate spiral ground state manifold. For that purpose, we perform a minimization of the eigenvalues in Eq.~(\ref{eq:Eval}) by requiring that the gradient $\nabla_\mu = \frac{\partial}{\partial q_\mu}$ ($\mu=x,y,z$) of the smallest eigenvalue must vanish,
\begin{equation}
\nabla \lambda_\nu(\q) = \nabla  J_{11}(\q) - \frac{1}{2|J_{12}(\bm{q})|} \nabla  |J_{12}(\q)|^2 = 0\;. \label{eq:GS}
\end{equation}
The diagonal matrix elements are of the form $J_2^\star f(\q)$, where $f(\q)$ is a real function containing a sum of cosine terms as in Eq.~(\ref{eq:fq}). The star in $J_2^\star$ (not to be confused with complex conjugation marked by an asterisk `$^*$') indicates that this is a generalized second-neighbor coupling which may also contain longer-range couplings within the same sublattice such that the sum in Eq.~(\ref{eq:fq}) also includes lattice vectors between further neighbors. In the next section we will make this generalization more explicit. The absolute value of the off-diagonal elements $|J_{12}(\q)|$ can be further specified using Eq.~(\ref{eq:Offdiagonal}),
\begin{equation}
  |J_{12}(\q)| = \frac{|J_1|}{2}\sqrt{\3\left( \sum_{\bm{R}^\text{nn}_{12}}{e^{i \q \bm{R}^\text{nn}_{12}}} \right) \3\3 \left(\sum_{\bm{R}^\text{nn}_{12}}{e^{i \q \bm{R}^\text{nn}_{12}}} \right)^*}.\label{eq:absJ12unconvenient}
\end{equation}
For nearest neighbors, each distance vector $\bm{R}^\text{nn}_{12}$ in this expression consists of a sum of lattice vectors $\bm{a}_i$ and the basis vector $\bm{h}$, where the exponentials $e^{i \q \bm{h}}$ cancel out when being multiplied by their complex conjugate. Therefore, Eq.~(\ref{eq:absJ12unconvenient}) can be recast in the more convenient form
\begin{equation}
    |J_{12}(\q)|  = \frac{|J_1|}{2} \sqrt{ z + p g(\q) }\;, \label{eq:absJ12}
\end{equation}
where $z$ is the coordination number of the lattice which results from contributions $e^{i \q \bm{R}^\text{nn}_{12}} e^{-i \q \bm{R}^\text{nn}_{12}} =1$. Furthermore, $p g(\q)$ is a sum of cosine terms of the form $\cos \left[\q  (\bm{a}_i - \bm{a}_j) \right]$ where the prefactor $p$ is defined such that the smallest coefficient of cosines in $g(\q)$ is one (see Sec.~\ref{ch:SSL} for explicit examples). Using Eq.~(\ref{eq:GS}), possible Luttinger-Tisza eigenvalues have to fulfill the condition
\begin{equation}
  J_2^\star \nabla f(\q) - \frac{|J_1|}{4} \frac{p}{\sqrt{ z + p g(\q) }} \nabla g(\q) = 0\;. \label{eq:GS2}
\end{equation}
In the generic case, this equation corresponds to one condition for each component of $\q$ such that Eq.~(\ref{eq:GS2}) is only fulfilled for a single spiral state. However, a special situation emerges when $f(\q) = g(\q)$, where the equation is either fulfilled by $\nabla f(\q) = 0$ or another set of solutions which is characterized by only {\it one} equation
\begin{equation}
  J_2^\star - \frac{|J_1|}{4} \frac{p}{\sqrt{ z + p f(\q) }} = 0\;,
\end{equation}
or equivalently
\begin{equation}
  f(\q) = p \left( \frac{J_1}{4 J_2^\star}\right)^2 - \frac{z}{p}\;. \label{eq:degSurf}
\end{equation}
Particularly, this single condition allows for a continuous manifold of solutions $\q$ forming closed contours (surfaces) for two (three) dimensional systems, hence, yielding spiral spin liquids. This is exactly the case for the $J_1$-$J_2$ model on the diamond and honeycomb lattices as studied in Refs.~\onlinecite{bergman07,mulder10}. For the bcc lattice \cite{attig17} the condition $ g(\q) = f(\q)$ requires the incorporation of third and fourth neighbor couplings $J_3$, $J_4$ into $J_2^\star$. All these models have in common that for $J_1<0$ ($J_1>0$) and $J_2^\star=0$ the system is first in a ferromagnetic (N\'eel) ground state, i.e., there is only one Luttinger-Tisza solution. However, when $J_2^\star/|J_1|$ is tuned beyond a certain critical ratio, Eq.~(\ref{eq:degSurf}) yields degenerate solutions. Since $f(\q)$ contains only cosine terms, its leading terms are quadratic in $q_{\mu}$. Therefore, when the spiral surface with small $|\q|$ just emerges from ferromagnetic or N\'eel order its shape is determined by the equation of an ellipsoid $\frac{q_x^2}{c_x}+\frac{q_y^2}{c_y}+\frac{q_z^2}{c_z} = 1$ or a sphere, as for the aforementioned models. While the above considerations hold for both signs of $J_1$, we will only treat the case of antiferromagnetic $J_1>0$ below.

It must, furthermore, be stressed that there are other ways in which Eq.~(\ref{eq:GS2}) can result in a degenerate manifold, e.g., if two or more equations are linearly dependent. In addition, the solutions to Eq.~(\ref{eq:degSurf}) do not pose a necessary condition for absolute minima of the eigenvalues, so it has to be checked for each individual problem that this is indeed the case.

\section{Further models with degenerate spiral surfaces}
\label{ch:SSL}
Based on the condition presented in Sec.~\ref{sec:Criteria}, it is now possible to construct new models that exhibit spiral spin-liquid phases. The general recipe which works for all bipartite lattices amounts to first expressing $|J_{12}(\q)|$ as in Eq.~(\ref{eq:absJ12}) and then finding the set of couplings $J_2^\star$ within the same sublattice such that $f(\q) = g(\q)$ is satisfied. Below we demonstrate this procedure for the square and the hcp lattice.

\subsection{Square lattice}\label{sub:square}
\begin{figure}[t]
         \includegraphics[width=0.99\linewidth]{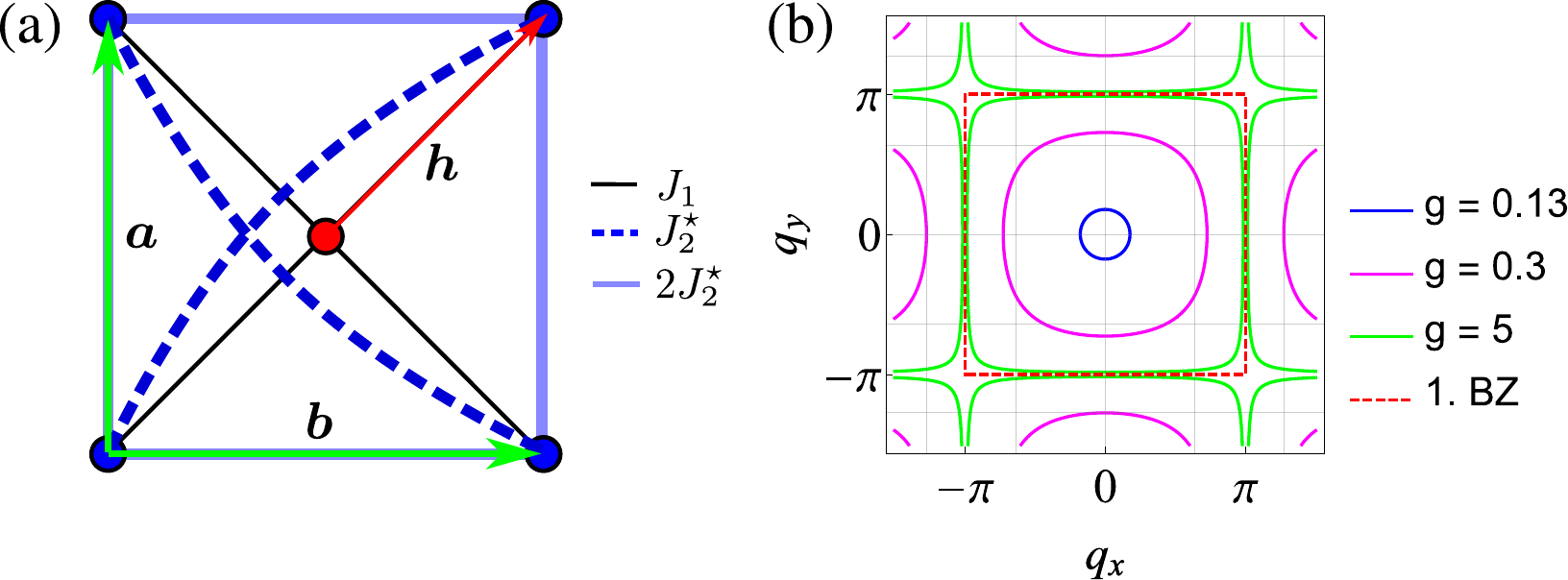}
  \caption{(a) Illustration of the square lattice when decomposed into two sublattices (blue and red dots). The lattice vectors are denoted by $\bm{a}$, $\bm{b}$ and the basis is given by the vector $\bm{h}$. For first, second, and third-neighbor couplings as indicated in the figure, the system establishes a manifold of spiral ground states. (b) Contours of degenerate ground states in momentum space for different coupling ratios $g=J_2^\star/J_1$ using the convention $|\bm{a}|=|\bm{b}|=1$. Note that N\'eel order corresponds to the $\Gamma$-point at $\q=0$.}\label{fig:CenteredSquare}
\end{figure}

The square lattice can be decomposed into two interpenetrating square lattices which are rotated by $45^\circ$ and stretched by a factor $\sqrt{2}$ as compared to the original one, see Fig.~\ref{fig:CenteredSquare}(a). The off-diagonal elements of $\underline{\bm{J}}(\q)$ take the simple form
\begin{equation}
  J_{12}(\q) = \frac{J_1}{2} e^{-i \q \bm{h}}\left(1 + e^{i \q \bm{a} } + e^{i \q \bm{b} }+e^{i \q( \bm{a}+ \bm{b}) }\right)\;,
\end{equation}
and we assume $J_1>0$. Because the square lattice is a Bravais lattice, $J_{12}(\q)$ is real. Using the notation $q_a \equiv \q \bm{a}$ and $q_b \equiv \q \bm{b}$ where $\bm{a}$ and $\bm{b}$ are the primitive lattice vectors, the absolute value $|J_{12}(\q)|$ can be written as
\begin{align}
   |J_{12}(\q)|&=\frac{J_1}{2} \{4+ 2[\cos(q_a - q_b) + \cos(q_a + q_b)\notag\\
   &\hspace*{1.9cm}+2 \cos(q_a) + 2 \cos(q_b)]\}^{\frac{1}{2}}\notag\\
   &\equiv \frac{J_1}{2} \sqrt{4+ 2 g(\q)}\;.\label{eq:ConditionSquare}
\end{align}
According to Eq.~(\ref{eq:fq}) the diagonal elements of $\underline{\bm{J}}(\q)$ have the form $J_{11} = J_2^\star f(\q)$, where $J_2^\star$ contains second and (possibly) further-neighbor couplings and $f(\q)$ is a sum of the cosine terms corresponding to the Fourier transforms of these couplings. The strengths of the further-neighbor couplings need to be adjusted such that the condition $f(\q) = g(\q)$ from Sec.~\ref{sec:Criteria} is satisfied. As can be seen from the function $g(\q)$ in square brackets of Eq.~(\ref{eq:ConditionSquare}), second-neighbor couplings $J_2$ [generating terms $\cos(q_{a/b})$] and third-neighbor couplings $J_3$ [generating terms $\cos(q_{a}\pm q_b)$] are sufficient to fulfill this condition. Furthermore, according to the prefactors of the cosine terms, the contributions from the second-neighbor couplings need to be twice as large as those from the third-neighbor couplings. In other words, $J_2^\star$ comprises a second-neighbor coupling $J_2$ of the strength $2 J_2^\star$ and a third-neighbor coupling $J_3$ of the strength $J_2^\star$, as shown in Fig.~\ref{fig:CenteredSquare}(a).

The degenerate ground states are spirals that satisfy Eq.~(\ref{eq:degSurf}) with $z = 4$ and $p = 2$, i.e.,
\begin{equation}
f(\q) = \frac{1}{8}\left(\frac{J_1}{J_2^\star}\right)^2 - 2\;.\label{eq:fqSquare}
\end{equation}
The evolution of the degenerate spiral contour as a function of the coupling ratio $g=J_2^\star/J_1$ is illustrated in Fig.~\ref{fig:CenteredSquare}(b). For $g\leq g_\text{c}=1/8$ there is no solution to Eq.~(\ref{eq:fqSquare}), i.e. the Luttinger-Tisza wave vector is determined by the condition $\nabla f(\bm{q})=0$ and the system resides in a N\'eel ordered phase. Above the critical value $g_\text{c}=1/8$ spiral contours begin to form out of the N\'eel order. Close to $g_\text{c}$ the contours first have a circular shape but with increasing $g$ they continuously deform into the square shape of the first Brillouin zone. In the limit $g\rightarrow\infty$ the two sublattices decouple where each sublattice individually forms a square lattice model with first and second-neighbor interactions where the latter are exactly half as strong as the former. This is the well-known case of the $J_1$-$J_2$ square lattice model at the classical transition point between N\'eel order and collinear order\cite{chandra88}. It is worth noting that the evolution of the spiral contours is very similar to the three-dimensional bcc lattice\cite{attig17}, where, likewise, the degenerate manifold adopts the shape of the first Brillouin zone in the limit $g\rightarrow\infty$. One may also embed the spiral system studied here into the classical phase diagram of the $J_1$-$J_2$-$J_3$ model with unrestricted couplings. The line cut in parameter space defined by $g$ then corresponds to a phase boundary between a $(q,q)$ spiral and a $(q,\pi)$ spiral, both of which do not exhibit any extensive degeneracies.\cite{moreo90,chubukov91}

\subsection{hcp lattice}\label{sub:hcp}
\begin{figure}[t]
\centering
\includegraphics[width=0.5\linewidth]{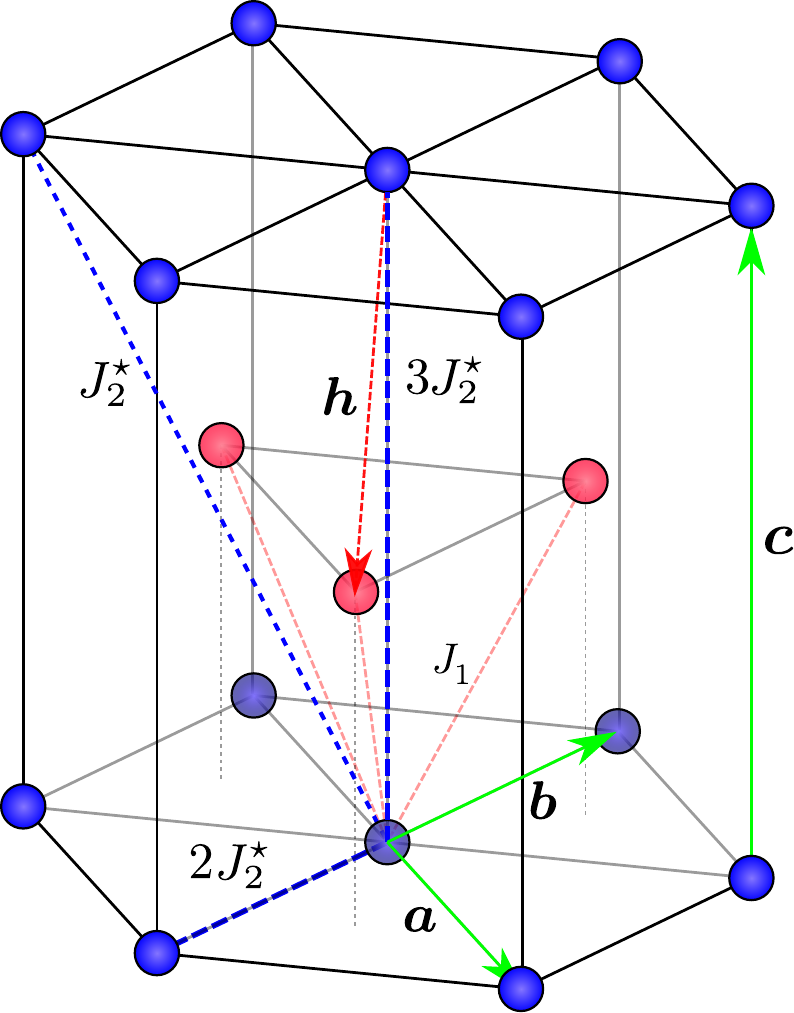}
\caption{The bipartite hcp lattice is built from an alternating sequence of stacked triangular lattices (blue and red) where the height of the unit cell is denoted by $c=|\bm{c}|$. The closest packing of equally sized spheres is realized for $c = \sqrt{8/3} \approx 1.633$ (in units of the in-plane nearest-neighbor distance). Spiral surfaces emerge if the longer range couplings $J_2^\star$ are given as illustrated.}
\label{fig:hcp}
\end{figure}
We finally discuss another spiral spin-liquid phase which is based on the three-dimensional hcp lattice. The hexagonal unit cell of the hcp lattice is spanned by the vectors
$\bm{a} = (\frac{1}{2},-\frac{\sqrt{3}}{2},0)$, $\bm{b}= (\frac{1}{2},\frac{\sqrt{3}}{2},0)$, $\bm{c}= (0,0,c)$ and the basis $\bm{h} = (0,-\frac{\sqrt{3}}{4},-\frac{c}{2})$. One may view the hcp lattice as an `abab' stacking of equilateral two-dimensional triangular lattices where the alternating sequence of these layers yields a bipartite geometry, see Fig.~\ref{fig:hcp}.
\begin{figure*}
    \centering
    \begin{minipage}[b]{0.25\linewidth}
        \includegraphics[width=\textwidth]{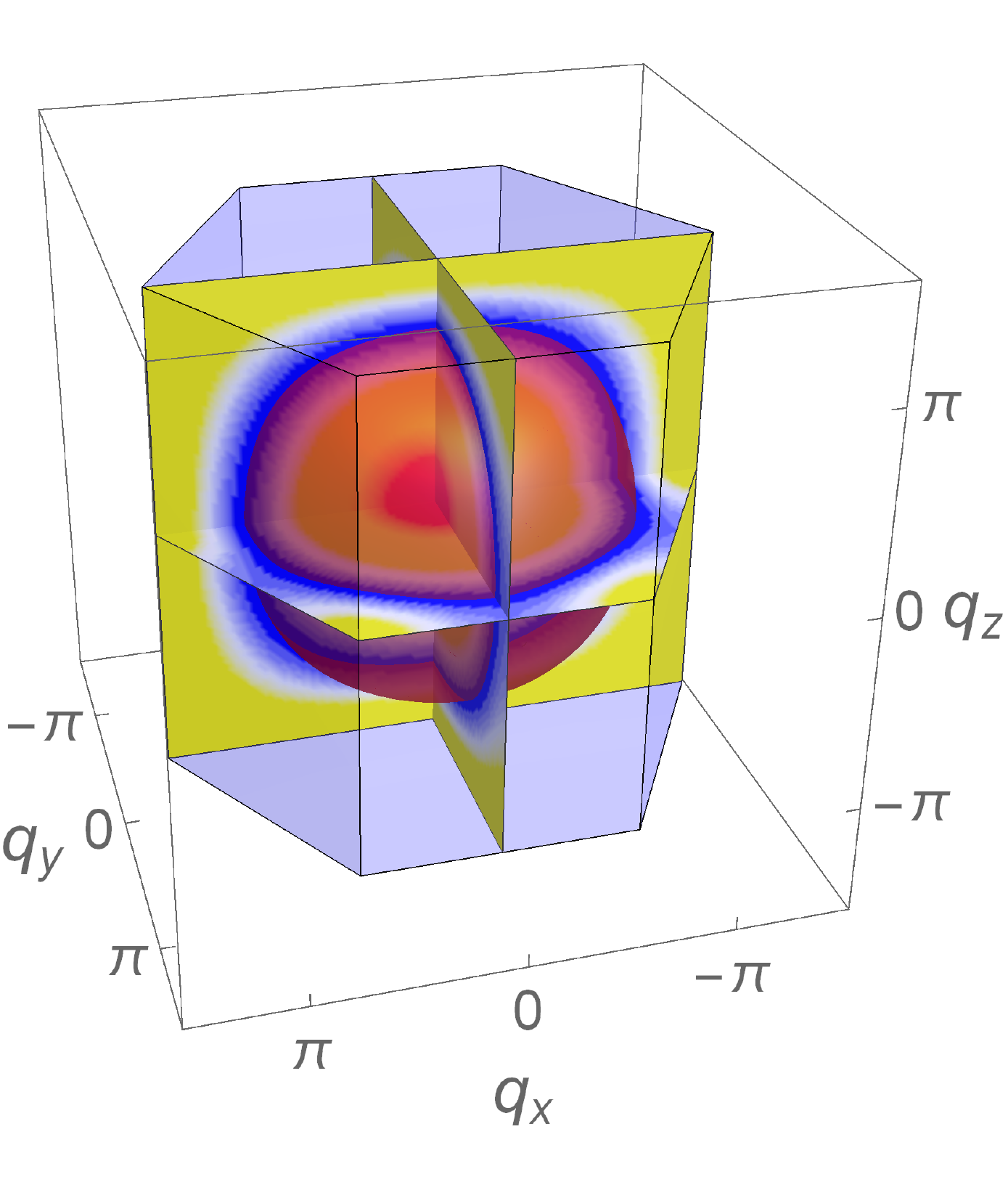}\\[-0.3cm]
        $g = 0.2$
    \end{minipage}
      ~
    \begin{minipage}[b]{0.25\linewidth}
        \includegraphics[width=\textwidth]{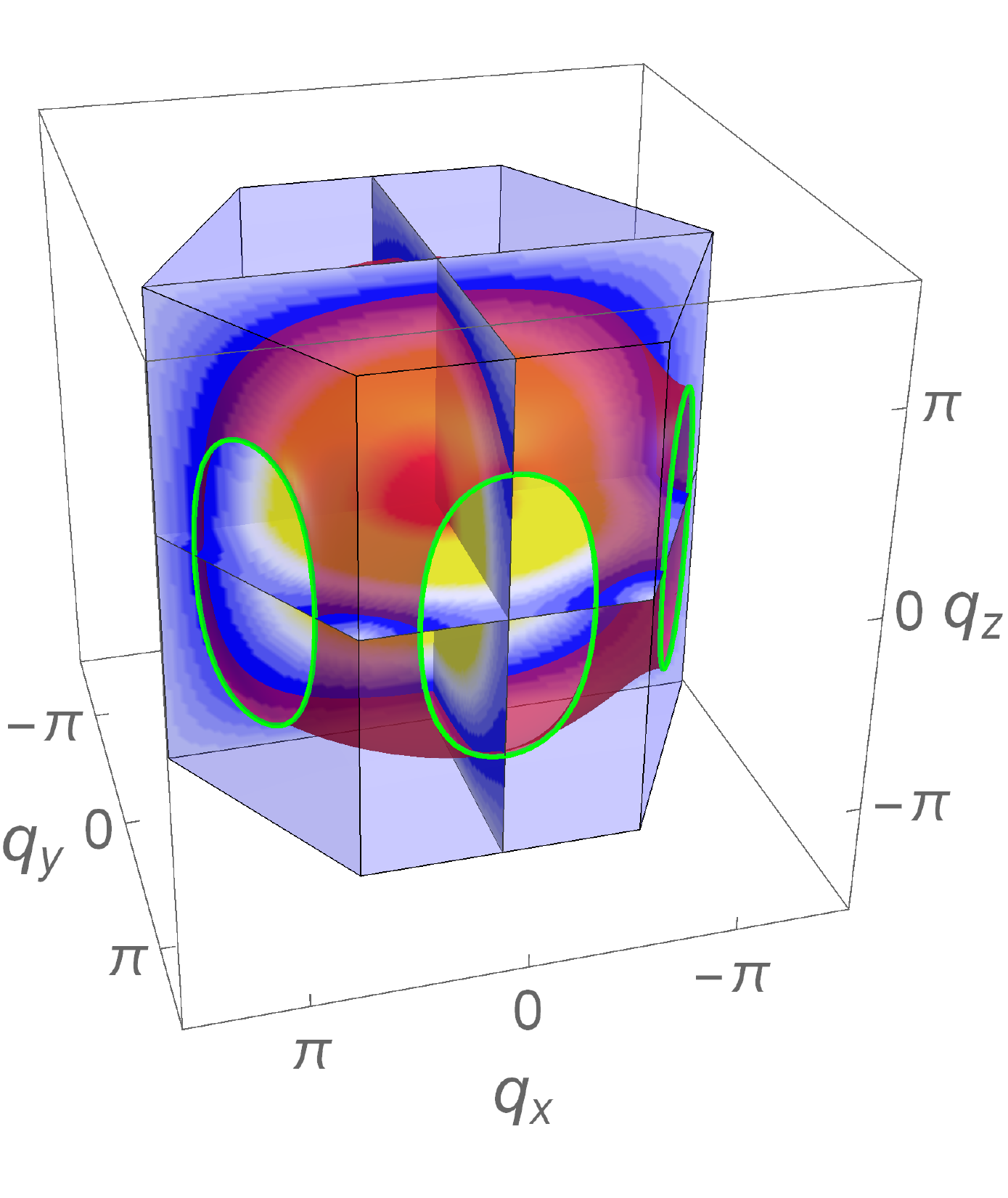}\\[-0.3cm]
        $g = 0.4$
    \end{minipage}
    ~
    \begin{minipage}[b]{0.25\linewidth}
        \includegraphics[width=\textwidth]{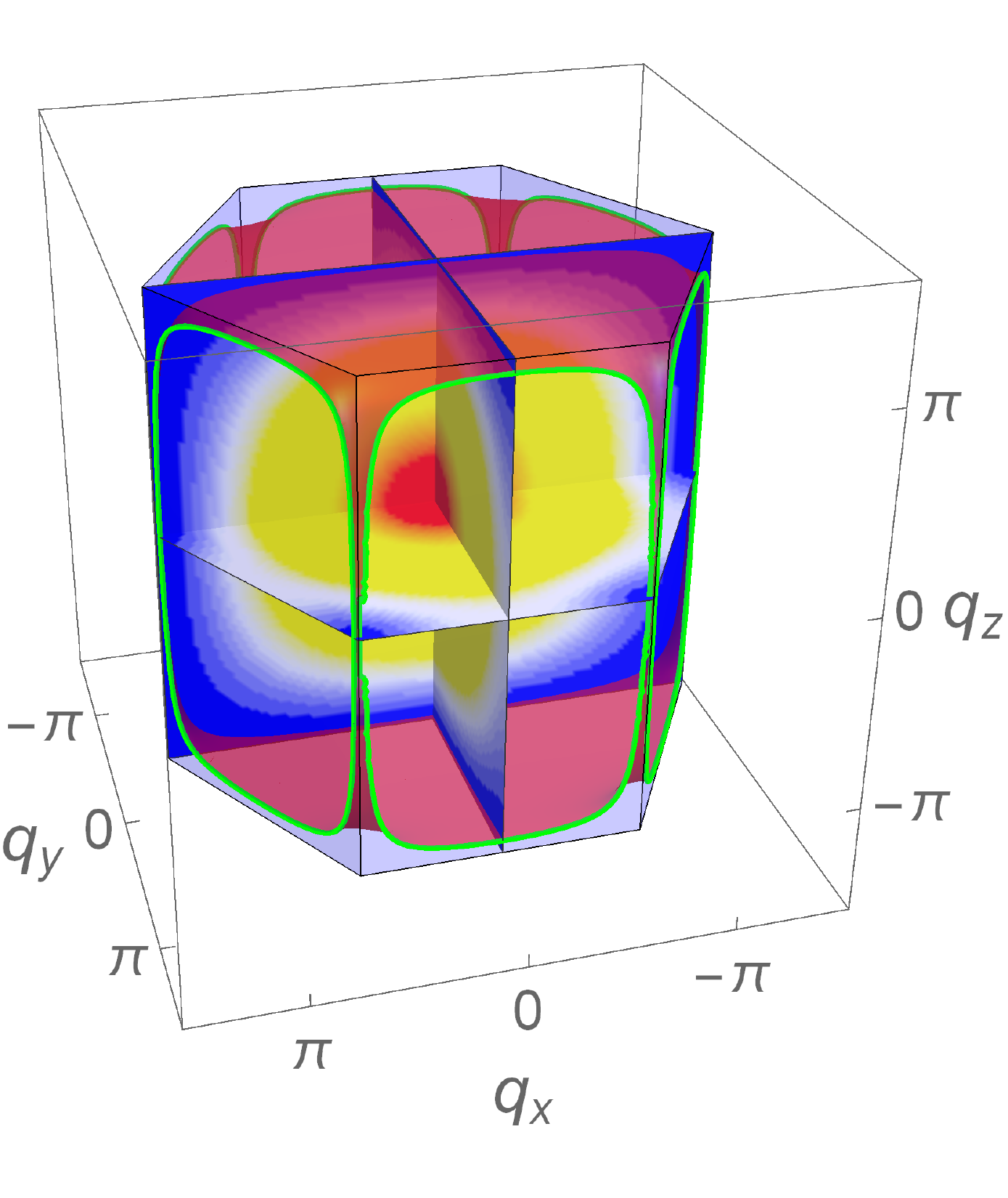}\\[-0.3cm]
        $g = 3$
    \end{minipage}
    ~

  \caption{Spiral surfaces on the hcp lattice for different coupling ratios $g=J_2^\star/J_1$. Displayed is the value of $\lambda_{\nu_{LT}}(\q)$ on several slices through reciprocal space, with blue being the lowest energy and red the highest. Further included are the ground-state spiral surfaces corresponding to the minima of $\lambda_{\nu_{LT}}(\q)$ (red) and their intersections with the first Brillouin zone (green). The plots have been obtained for $c= \sqrt{2/3}$.\label{fig:hcp_surfaces}}
\end{figure*}

The off-diagonal element of the coupling matrix $\underline{\bm{J}}(\q)$ is given by
\begin{align}
J_{12}(\q) &= \frac{J_1}{2} e^{i \bm{q} \bm{h}} \left[1 + e^{-i \bm{q} \bm{a}} + e^{i \bm{q} \bm{b}}
+ e^{i \bm{q} \bm{c}} + e^{i \bm{q} (\bm{c} + \bm{b})}\right.\notag\\
&\left. +\; e^{-i \bm{q} (\bm{c} - \bm{a})} \right]
\end{align}
and its absolute value has the form
\begin{equation}
  |J_{12}(\q)| = \frac{J_1}{2} \sqrt{6 + 2 g(\q)}
\end{equation}
with
  \begin{align}
  g(\q) &= 2[\cos{q_a}+\cos{q_b}+\cos{(q_a + q_b)}] + 3\cos{q_c} \nonumber\\
  &+ \cos{(q_a+q_c)}+ \cos{(q_c-q_b)} + \cos{(q_b+q_c)}\nonumber\\
  &+ \cos{(q_a-q_c)}+ \cos{(q_b-q_c)}+ \cos{(q_a+q_b+q_c)}\;,
  \end{align}
where we set $J_1$ to be antiferromagnetic and $q_a=\bm{q} \bm{a}$, $q_b=\bm{q} \bm{b}$, $q_c=\bm{q} \bm{c}$. The condition $f(\q) = g(\q)$ again determines the set of couplings $J_2^\star$ which is needed to generate a spiral degeneracy. By simple bookkeeping of terms one finds that one in-plane coupling of size $2J_2^\star$ as well as two couplings $J_2^\star$ and $3 J_2^\star$ connecting triangular layers separated by the distance $c$ are required, see Fig.~\ref{fig:hcp}. The spiral surface then contains all spiral states with wave vectors $\q_{LT}$ that satisfy the equation
\begin{equation}
  f(\q) = \frac{1}{8} \left(\frac{J_1}{J_2^\star} \right)^2 -3\;. \label{eq:hcp}
\end{equation}
As before, as a function of the coupling ratio $g=J_2^\star/J_1$ the system first shows N\'eel order. Above the critical coupling $g_\text{c} = 1/12$ the condition in Eq.~(\ref{eq:hcp}) has finite solutions such that spiral surfaces appear. The shape of the spiral surface emerging for $g>g_\text{c}$ also depends on the length $c=|\bm{c}|$. Interestingly, for the closest packing $c= \sqrt{8/3}$, it is not a sphere but an ellipsoid. Changing the length $c$ (which is equivalent to rescaling $q_c$) results in a stretching/compression of the surface in the vertical direction. By expanding $f(\q)$ in Eq.~(\ref{eq:hcp}) up to second order in $\bm{q}$ one finds that for $c = \sqrt{2/3}$ the surface for $g$ just above $g_c$ is exactly spherical. In Fig.~\ref{fig:hcp_surfaces} we plot the degenerate spiral manifolds for various values of $g$. As can be seen, for $g > 0.25$ the surface cuts through the edges of the first Brillouin zone and in the limit $g \rightarrow \infty$ the surface only consists of flat planes at $\q_z = \pm \pi$ connected by nodal lines along the vertical edges of the first Brillouin zone.

\section{Effect of quantum fluctuations}
\label{sec:QSL}

In this section, we discuss the fate of the spiral spin liquids identified in the last section when quantum fluctuations are included, i.e., when tuning the magnitude $S$ of the spins away from the classical limit. For this purpose, we employ the pseudofermion functional renormalization group (PFFRG) method for spin systems in its implementation for unrestricted $S$.\cite{reuther10,baez18} Particularly, we calculate the magnetic correlations for the square and hcp lattice models as a function of the coupling ratio $g$ and the spin length $S$ and determine whether or not the systems develop magnetic long-range order. The results discussed below are based on the imaginary-time one-loop plus Katanin PFFRG approach for spin systems at $T=0$ as described in Ref.~\onlinecite{baez18}. The method uses an Abrikosov decomposition of each spin operator into two pseudofermions~\cite{abrikosov65,abrikosov70} which allows for an application of fermionic diagram techniques. The purely quartic fermionic Hamiltonians which arise after the Abrikosov decomposition of the Heisenberg interactions are notoriously difficult to solve. The PFFRG handles this problem by first introducing an artificial cutoff $\Lambda$ (the so-called RG scale) which sets the bare fermionic single-particle Green's function to zero for Matsubara frequencies $|\omega|<\Lambda$. With this manipulation, the system's $n$-particle vertex functions can be calculated from an infinite set of coupled integro-differential equations\cite{wetterich93} which needs to be truncated for numerical solubility. Here, we use a well-established truncation scheme which takes into account one-loop and Katanin-type diagrammatic contributions.\cite{katanin04} The resulting set of coupled equations is solved with the initial conditions of the vertex functions at $\Lambda \rightarrow \infty$ given by the bare interactions. Physical vertex functions are then obtained in the limit $\Lambda \rightarrow 0$ by successively integrating out energy degrees of freedom. In essence, the PFFRG performs diagrammatic resummations which systematically incorporate the large-$N$ and large-$S$ limits [$N$ determines the spins' symmetry group $SU(N)$]. In particular, it has been proven that the leading diagrammatic contributions in $1/N$~\cite{buessen18_2,roscher18} and in $1/S$ are both exactly summed up, where in the large $S$ limit the PFFRG becomes identical to the Luttinger-Tisza method.\cite{baez18} For more details on the method, please refer to Refs.~\onlinecite{reuther10,baez18}.

We use the PFFRG to calculate the system's isotropic static spin correlations
\begin{equation}\label{eq:StaticSus}
\chi_{ij}(\Lambda)\equiv\chi_{ij}^{zz}(\Lambda)=\int_0^\infty d\tau\langle\hat{S}_i^z(\tau)\hat{S}_j^z(0)\rangle_\Lambda\;,
\end{equation}
where $\hat{S}_i^\mu(\tau)=e^{\tau\hat{H}}\hat{S}_i^\mu e^{-\tau\hat{H}}$ and the bracket $\langle \dots \rangle_{\Lambda}$ denotes that the expectation value is computed at the RG scale $\Lambda$. Fourier-transforming $\chi_{ij}$ into momentum space, a smooth flow of the maximal $\bm{q}$-component of the susceptibility towards the physical limit $\Lambda\rightarrow 0$ indicates that the system is in a paramagnetic phase, hence, possibly realizing a quantum spin liquid. The onset of magnetic order, on the other hand, is accompanied by a divergence or a kink in the susceptibility flow. These two distinct behaviors are explained by the fact that our PFFRG formalism is invariant under a global $SU(2)$ spin-rotation. Once this symmetry is spontaneously broken due to the onset of magnetic order, the algorithm is unable to correctly describe the system's correlations and the RG flow becomes unphysical. A true divergence at a finite critical RG scale $\Lambda_{\mathrm{c}}$ would be expected if correlations between infinitely separated lattice sites and continuous Matsubara frequencies were included. Since we use discretized frequencies and truncate the extent of the spin correlations in real space, such divergences are typically regularized to a kink in the susceptibility flow. Hence, in the case of a paramagnetic phase we can compute the ${\q}$-dependent susceptibility down to the physical limit $\Lambda\rightarrow0$. Alternatively, from a kink or a divergence in the susceptibility flow we can infer that the system is magnetically ordered where the specific type of order is determined by the position of the magnetic Bragg peaks in $\bm{q}$ space. For the models considered here, we use a mesh of $70$ discretized points for all three Matsubara frequency arguments of the two-particle vertex.

It is worth highlighting that in the absence of an instability feature during the RG flow, our analysis below only allows us to conclude that the system is non-magnetic but does not yet imply a quantum spin-liquid phase. This is because a non-magnetic ground state may still exhibit some type of hidden order such as, e.g., dimer crystal formation. Extended PFFRG schemes\cite{reuther10,iqbal19,iqbal16_2} (which will not be applied here) also allow one to probe the system with respect to dimer order. If such type of order is absent as well, a recently developed PFFRG approach is applicable which has shown some success in determining the system's low-energy spinon band structure of the putative quantum spin liquid.\cite{hering19}
\begin{figure*}[t]
\centering
\includegraphics[width=0.99\textwidth]{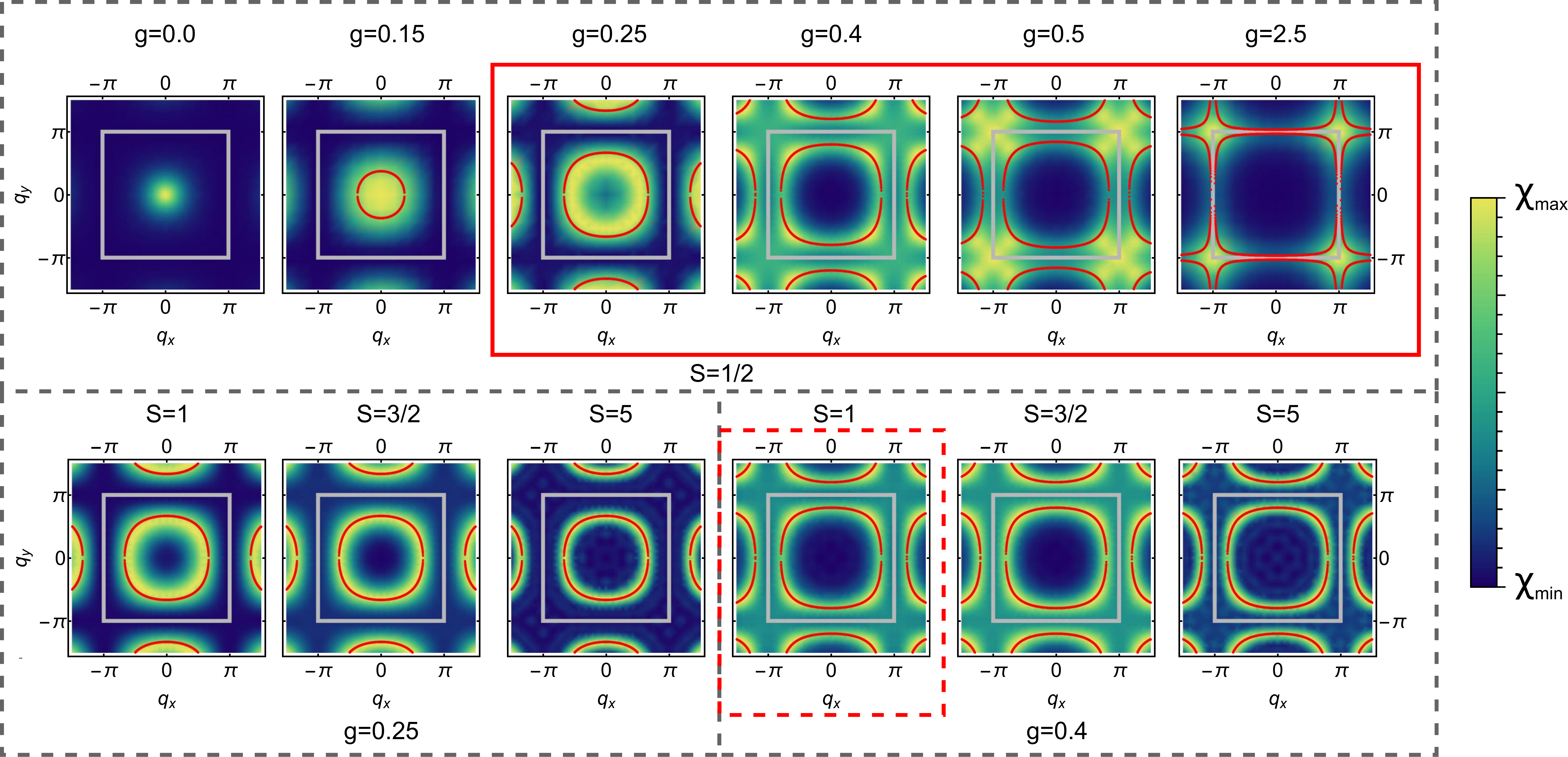}
\caption{Static spin susceptibilities $\chi(\bm{q})$ for the square lattice Heisenberg model with couplings $g=J_2^\star/J_1$ as determined in Sec.~\ref{sub:square}. The top panel shows $\chi(\bm{q})$ for varying $g$ and fixed $S=1/2$ while the lower panels use $g=0.25$ and $g=0.4$, respectively, and vary $S$. Red lines in the plots depict the classical spiral contours identified by the Luttinger-Tisza approach in Sec.~\ref{sub:square}. Gray lines illustrate the boundary of the first Brillouin zone. Full red frames around the plots indicate a paramagnetic phase and dashed red frames signal parameter regimes of uncertain flow behavior. In non-magnetic phases the susceptibility is plotted in the limit $\Lambda \rightarrow 0$ while otherwise the plots are shown at the corresponding critical RG scale $\Lambda_{\mathrm{c}}$ (cf. main text). \label{fig:SquareSusComp}}
\end{figure*}

\subsection{Square lattice}
\label{sec:SquareFRG}
We now present the results of our PFFRG calculations for the square lattice Heisenberg model with couplings as derived in Sec.~\ref{sub:square}. The lattice conventions are the same as above, i.e., the primitive lattice vectors $\bm{a}$, $\bm{b}$ are defined as in Fig.~\ref{fig:CenteredSquare} and have unit length $|\bm{a}|=|\bm{b}|=1$. While the PFFRG, in principle, treats an infinite lattice, finite-size effects still enter since spin correlations are only taken into account up to $10$ nearest-neighbor distances in both spatial directions and are treated as zero beyond this length. This means that, in total, each spin is coupled to $440$ surrounding spins. The free parameters of our simulations are $g=J_2^\star/J_1$ and $S=1/2, \, 1, \, 3/2,\, \dots$.

In Fig.~\ref{fig:SquareSusComp} we show the momentum-resolved susceptibility $\chi(\bm{q})$ for varying $g$ and fixed $S=1/2$ as well as for fixed $g=0.25$ and $g=0.4$ and varying $S$. For a better comparison with the results from the previous sections, the Fourier transforms from $\chi_{ij}$ to $\chi(\bm{q})$ are performed with respect to one sublattice only, i.e., the Fourier sums only run over $i$, $j$ which belong to the same sublattice. As a consequence, antiferromagnetic N\'eel order manifests in a magnetic Bragg peak at $\bm{q}=0$. Explicitly taking into account both sublattices would result in an additional modulation of the susceptibility with a function that is periodic in the second Brillouin zone. This however, would only complicate the comparison with the classical Luttinger-Tisza results but would not yield relevant new insights.
\begin{figure*}
 \includegraphics[width=0.7\textwidth]{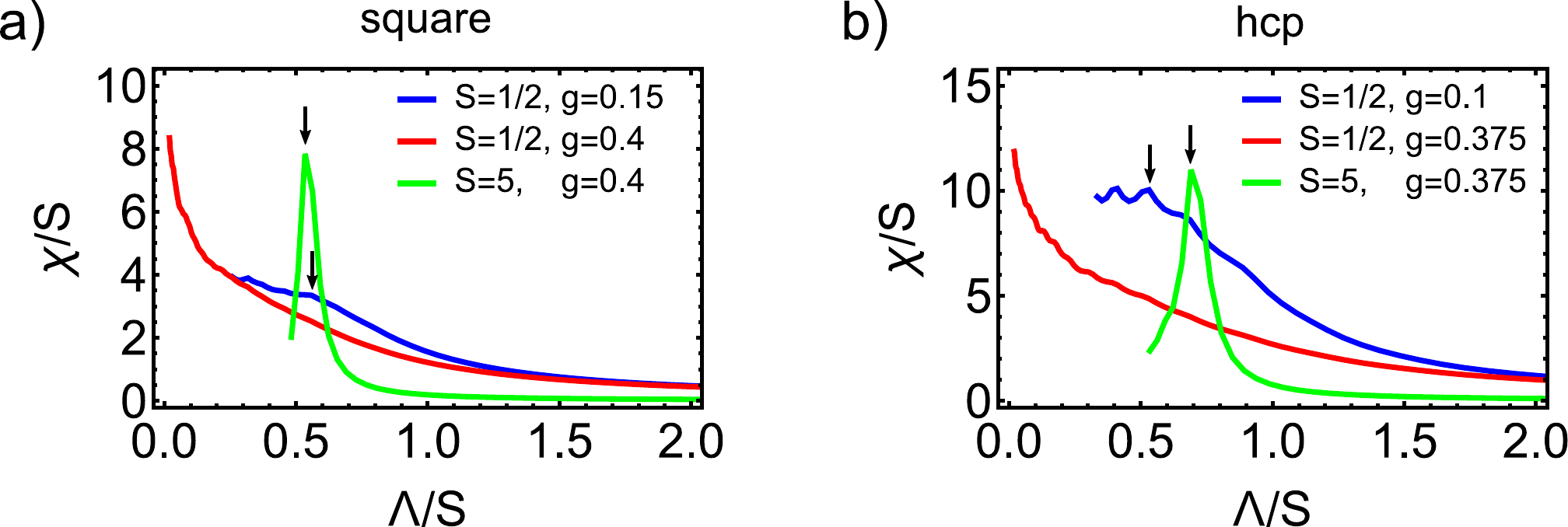}
  \caption{Flows of the maximal spin susceptibility in momentum space as a function of the RG scale $\Lambda$ at selected coupling ratios $g$ and spin magnitudes $S$ for (a) the square lattice model and (b) the hcp lattice model. The susceptibility $\chi$ and the RG scale $\Lambda$ are rescaled by the spin magnitude $S$ for better comparisons of the flows. A kink or cusp marked by an arrow indicates a magnetic instability whereas a smooth flow towards $\Lambda\rightarrow 0$ suggests that the system is non-magnetic.
 }\label{fig:SquareFlows}
\end{figure*}
\begin{figure*}[t]
\centering
\includegraphics[width=0.99\textwidth]{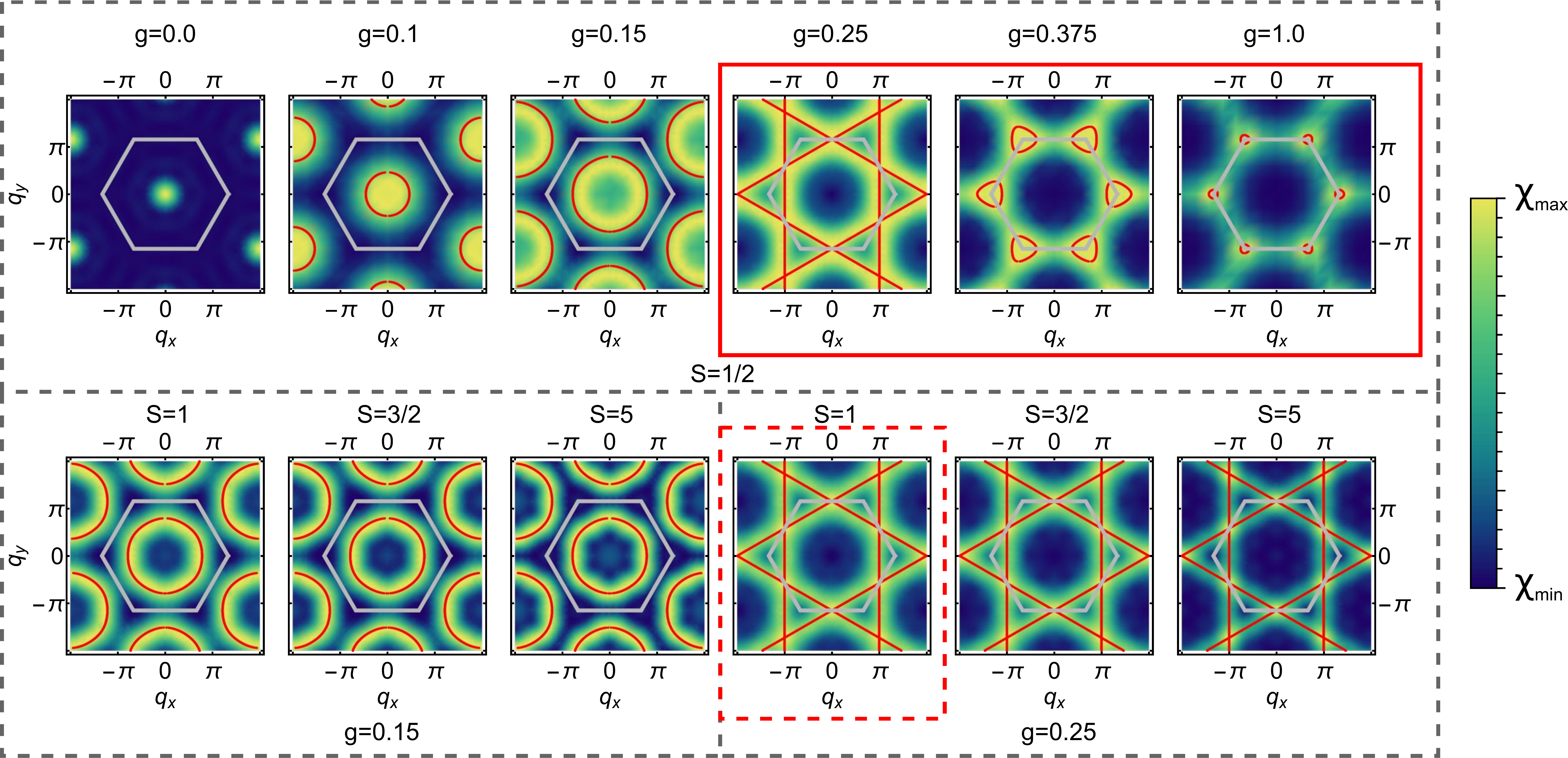}
\caption{Static spin susceptibilities $\chi(\bm{q})$ in the $q_z=0$ plane for the hcp lattice Heisenberg model with couplings $g=J_2^\star/J_1$ as derived in Sec.~\ref{sub:hcp}. The top panel contains selected susceptibility plots for fixed $S=1/2$ and varying $g$ while the two lower panels use fixed ratios $g=0.15$ and $g=0.25$ and vary $S$. Red lines in the plots illustrate the classical spiral contours identified by the Luttinger-Tisza approach in Sec.~\ref{sub:hcp}. Gray lines mark the boundary of the first Brillouin zone. Full red frames around the plots indicate non-magnetic behavior and dashed red frames signal parameter regimes of uncertain flow behavior. In non-magnetic phases the susceptibility is plotted in the limit $\Lambda \rightarrow 0$ while otherwise the plots are shown at the corresponding critical RG scale $\Lambda_{\mathrm{c}}$. \label{fig:hcpSusComp}}
\end{figure*}

The susceptibility plots in Fig.~\ref{fig:SquareSusComp} demonstrate that even in the quantum limit $S=1/2$ the momentum distribution of the response still roughly follows the classical spiral contours. The characteristic ring-like pattern, though, sets in at slightly larger $g$ as compared to the classical limit, such that there is a small regime around $g=0.15$ where the classical system has a spiral degeneracy but the quantum system still shows a single (but broadened) peak at $\bm{q}=0$. From the flow of the susceptibility we infer that for $S=1/2$ a paramagnetic phase is realized for $g\gtrsim 0.2$ and that the system orders magnetically for smaller values of $g$ [see Fig.~\ref{fig:SquareFlows}(a) which illustrates $\Lambda$-dependent susceptibilities for selected $g$ and $S$ indicating either a magnetically ordered or a non-magnetic RG flow]. Particularly, the onset of the non-magnetic phase is seen to coincide with the appearance of ring-like features in the susceptibility, revealing a close connection between both phenomena. The system remains non-magnetic in the whole parameter regime up to $g\rightarrow\infty$ where the spins on the two sublattices decouple, each realizing a $J_1$-$J_2$ square lattice model with $J_2=0.5J_1$. Previous PFFRG studies found a similar extent of the non-magnetic phase at $S=1/2$\cite{reuther11_2} (including the limit $g\rightarrow\infty$\cite{reuther10}), however, their focus was not on spiral properties. Furthermore, the observation that quantum fluctuations enlarge the regime of antiferromagnetic N\'eel correlations at the expense of spiral correlations has already been made for other systems such as Heisenberg models on the honeycomb or bcc lattices.\cite{reuther11,ghosh19_2}

A closer inspection of the susceptibility profiles for $S=1/2$ in Fig.~\ref{fig:SquareSusComp} indicates some distinct differences as compared to the classical spiral contours. For $g\lesssim 0.25$ the rims in the susceptibility are found to be contracted towards the $\bm{q}=0$ point while for $g\gtrsim0.5$ the weight is shifted more towards $\bm{q}=(\pi,\pi)$. Furthermore, the contours of strong response show a selection of dominant wave vectors due to quantum fluctuations. These effects are most pronounced around $g=0.4$ where incommensurate momenta of the form $\bm{q}=(\pm q,\pm q)$ are selected and for $g\rightarrow\infty$ where the point $\bm{q}=(\pi,\pi)$ is preferred.

Increasing the spin magnitude, there is a small range of couplings $g\simeq0.3,\,\dots,\,0.5$ for $S=1$ where the system possibly still resides in a paramagnetic phase (the behavior of the RG flow in this regime is, though, less conclusive as compared to $S=1/2$). For all other parameter values and for higher spins, our calculations indicate a magnetically ordered ground state. As shown in the lower panels of Fig.~\ref{fig:SquareSusComp}, both quantum effects, i.e., contraction of spiral contours and selection of dominant wave vectors, are suppressed for increasing $S$ and the exact Luttinger-Tisza result is reproduced correctly as expected.~\cite{baez18}

\subsection{hcp lattice}
\label{sec:hcpFRG}
Next, we investigate the hcp lattice model via PFFRG. The coupling constants ranging up to fourth neighbors are defined in Sec.~\ref{sub:hcp} and our tuning parameters are again $g=J_2^\star/J_1$ and the spin magnitude $S$. Our calculations take into account spin correlations with a maximal length of $8$ nearest-neighbor distances such that each spin is coupled to $932$ surrounding spins. The momentum-resolved susceptibilities $\chi(\bm{q})$ are shown in Fig.~\ref{fig:hcpSusComp} for varying $g$ at $S=1/2$ (top panel) and varying $S$ for $g=0.15$ and $g=0.25$ (two bottom panels). We only plot $\chi(\bm{q})$ in the plane with $q_z=0$ which contains most of the quantum effects we wish to discuss. Furthermore, as in Sec.~\ref{sec:SquareFRG}, the susceptibility has been obtained by restricting to one sublattice only.

Despite the different lattice geometry and exchange couplings of this system, our observations are similar to those for the square lattice model in the previous subsection, i.e., there is an overall good agreement between the classical spiral surfaces and the regions of strong response in the calculated susceptibility. Starting with $S=1/2$, the onset of a ring-like susceptibility again occurs at a somewhat larger ratio $g$ than in the classical case. The flow of the susceptibility, which we plot in Fig.~\ref{fig:SquareFlows}(b) for selected parameter values, implies that there is an extended range $g\simeq 0.2, \, \dots,\,1$ where the spin-$1/2$ system is in a paramagnetic phase. Interestingly, in contrast to the square lattice model of the previous subsection, here we also identify a small parameter regime around $g=0.15$ where spiral surfaces are clearly visible in our PFFRG results but the system is still magnetically ordered. Within the paramagnetic phase the spiral contours in the $q_z=0$ plane first show a pattern of kagome-like streaks at $g=0.25$ which then contract towards the corners of the first Brillouin zone. We again observe small deviations between the classical degeneracies and the susceptibility distribution in the quantum case. Particularly, for small $g$ (see, e.g., $g=0.15$) the rings in the susceptibility are smaller than classically expected while for larger $g\gtrsim0.375$ they quickly become pinned to the corners of the first Brillouin zone. Selection effects at $S=1/2$ seem less pronounced compared to the square lattice model, nevertheless, a certain selection takes place outside the $q_z=0$ plane which is not visible in our plots. In the limit $g\rightarrow\infty$, the hcp layers decouple into an `aaa' stacking of triangular layers with weak indication of magnetic order.

With increasing $S$ the non-magnetic phase quickly shrinks, e.g., at $S=1$, there only exists a small regime $g\simeq 0.2, \, \dots,\,0.4$ where the system is possibly non-magnetic. For $S>1$ only magnetic phases remain and the susceptibility distribution shows increasingly better agreement with the Luttinger-Tisza result. However, we also emphasize that three-dimensional systems with large spin magnitudes as studied here are rather prone to finite-size effects within the PFFRG. This is because spin correlations have a strong tendency to become long-range when quantum fluctuations die out for $S\rightarrow\infty$. Furthermore, for three-dimensional systems we have to reduce the distance of the longest spin correlations in our algorithm to become numerically feasible. As a consequence, there may easily arise a situation where the correlation length is much larger than the distance of the longest correlations taken into account. This effect is most pronounced in our susceptibility plot for $g=0.25$ and $S=5$ showing a spurious selection of wave vectors at the midpoint positions of the edges of the Brillouin zone which are not expected at large $S$.

\section{Conclusions}\label{sec:conclusion}
In summary, this work investigates the capability of classical spiral spin liquids to induce quantum spin-liquid behavior when the spin magnitude is decreased down to $S=1/2$. For several spiral spin-liquid systems such as classical Heisenberg models on the honeycomb, diamond, and bcc lattices the onset of a non-magnetic quantum phase at $S=1/2$ has already been proposed in various numerical studies.\cite{buessen18,reuther11,zhu13,gong13,zhang13,li12,ghosh19_2} However, due to the lack of a common principle behind the precise arrangements of couplings in these systems, the family of classical spin models yielding spiral degeneracies has been very small so far. To overcome this deficiency, we have first developed a general and exact procedure to construct new spiral spin-liquid phases on bipartite lattices. As an example, we have demonstrated this approach based on the two-dimensional square lattice and the three-dimensional hcp lattice, both of which exhibit a one-parameter family of exchange couplings where continuous spiral degeneracies exist. We have further studied the impact of quantum fluctuations on these systems by employing the PFFRG method. We find that large portions of the classical spiral spin-liquid regime become non-magnetic quantum phases when the spin length is reduced to $S=1/2$. However, even in this extreme quantum limit spiral correlations still determine the system's magnetic properties on short length scales, as evidenced by pronounced rims of strong response at approximately the same $\bm{q}$-space locations as the classical spiral contours. Depending on the precise coupling ratios, the susceptibility is not evenly distributed along these rims but shows small maxima which indicate a quantum order-by-disorder effect.

While the spin models investigated in this work are interesting candidate systems for stabilizing exotic non-magnetic quantum phases, it needs to be emphasized again that based on our current analysis we cannot draw any definite conclusion about whether or not they indeed realize a quantum spin liquid. A well-known alternative is the occurrence of spontaneous dimer ordering which may also be checked within a modified PFFRG approach.\cite{reuther10,iqbal19,iqbal16_2} Such an analysis, however, is beyond the scope of the current work. Eventually, it would be desirable to realize the above lattice geometries and sets of exchange couplings in a real material. As a complicating fact, though, both models exhibit longer-range exchange couplings beyond second neighbors which need to appear in fixed ratios to each other. While it might be difficult to exactly realize these ratios in a material, our procedure for designing spiral spin liquids is very general and may be applied to all bipartite lattices. Particularly, as a future direction of research, it would be interesting to search for new spiral spin-liquid models more systematically and identify simpler and more realistic arrangements of exchange couplings.

\section*{Acknowledgements}
We would like to thank Stef Koenis for his feedback and helpful discussions. The PFFRG simulations were performed on the tron cluster at Freie Universit\"{a}t Berlin. This work was partially supported by the German Research Foundation within the CRC 183 (project A02).

\end{document}